\documentclass[11pt,a4paper,english]{article}
\usepackage[latin9]{inputenc}
\usepackage{float}
\usepackage{amsmath}
\usepackage{graphicx}
\usepackage{esint}
\makeatletter

\newcommand{\lyxdot}{.}

\pdfoutput=1 

\usepackage{jinstpub}

\title{\boldmath Numerical study of space charge electric field inside Resistive Plate Chamber}

\author[a,c,1]{Tanay Dey,\note{Corresponding author.}}
\author[a,b]{Supratik Mukhopadhyay,}
\author[c]{Subhasis Chattopadhyay,}
\author[c]{Jhilam Sadukan}

\affiliation[a]{Homi Bhabha National Institute,\\Mumbai, India}
\affiliation[b]{Saha Institute of Nuclear Physics,\\Kolkata,India}
\affiliation[c]{Variable Energy Cyclotron centre,\\Kolkata, India}

\emailAdd{tanay.jop@gmail.com}

\abstract{Resistive plate chamber (RPC) is one of the state-of-the-art particle detection technology for the High Energy Physics (HEP) experiments. The basic operating mechanism of an RPC involves ionization of gas due to the passage of charged particles followed by electron transport, avalanche, and subsequent electromagnetic induction on readout strips due to the movement of the electrons and ions. Especially during streamer mode of operation, the electric field applied to the RPC can get significantly modified due to the presence of a large number of electrons and ions.
In this study, we have worked on dominant issues related to the estimation of the electric field due to the space charge arising out of the presence of electrons, ions within an RPC. For this purpose, we have considered two approaches: representation of the space charge cloud as (a) a collection of ring charges, and (b) as a collection of line charges. The results from these different methods have been compared with the results available in the literature.}


\arxivnumber{} 

\collaboration[c]{}

\proceeding{XV$^{\text{th}}$ Workshop on Resistive Plate Chambers and Related Detectors - RPC2020\\
  10-14 February, 2020\\
   Rome, Italy}

\@ifundefined{showcaptionsetup}{}{%
 \PassOptionsToPackage{caption=false}{subfig}}
\usepackage{subfig}
\makeatother

\usepackage{babel}
\begin{document}
\maketitle \flushbottom

\section{Introduction}

Resistive plate chamber (RPC)\cite{cardeli-1,cardeli-2} is a popular
gaseous detector used to detect charged particles. It is basically made of two parallel
resistive bakelite or glass plates. The space between the electrodes contained
a particular gas mixture and a high-voltage (say $\pm5kV$) is applied at each plate. When high energy charged particle pass
through the RPC, it can knock out some primary electrons from gas
molecules along its path. As an effect of the high electric field
inside the gas gap of RPC, those primaries are accelerated towards
the electrodes and produce secondaries by the ionization process.
This process continues and develops an avalanche of numerous electrons
and ions. An signal pulse is induced due to the movement of electrons in a pick-up strip placed on the detector \cite{Sauli:2014cyf}. So It is clear that to investigate the detector physics
of RPC, we need to simulate that avalanche process very precisely.
We know that space charge plays a crucial role while an avalanche
is developing. So for simulation of an avalanche, the electric field
due to space charge needs to be calculated dynamically.

In this paper, we intend to discuss three different methods (A, B, C) to
calculate the electric field in the presence of the space charge.
The method-A and C (see section \ref{subsubsec:ring-approximation} and \ref{subsec:Method-available-in})
contains modeling of the space charge region as several co-centric
charged rings of gradually increasing radius as in ref. \cite{Lippmann_1}.
Again, a ring can be thought of as a collection of charged straight
lines (see section \ref{subsec:Straight-line-approximation-with})
of equal length $S$, which is our method-B. Now two cases need
to be discussed:

Case-1: the linear charge density ( $\lambda$ ) of a ring is kept
constant in methods A and C. The lines corresponding to a ring have been
shared the equal amount of charge ( $\lambda S$ ) to the method-B.

Case-2: the condition for the rings is remaining identical, as in
case-1. However, the charges of lines have been calculated separately.
We then compare the electric fields for these lines and rings in two
cases. In this initial phase we are ignoring the reflections of charges on the ground plates. This is a serious matter and will be taken up in subsequent work.

\section{Calculation of positions of electron and ion cluster }

\label{sec:position}
\begin{figure}
\centering\subfloat[\label{fig:position of elc clouds}]{\includegraphics[width=0.40\textwidth]{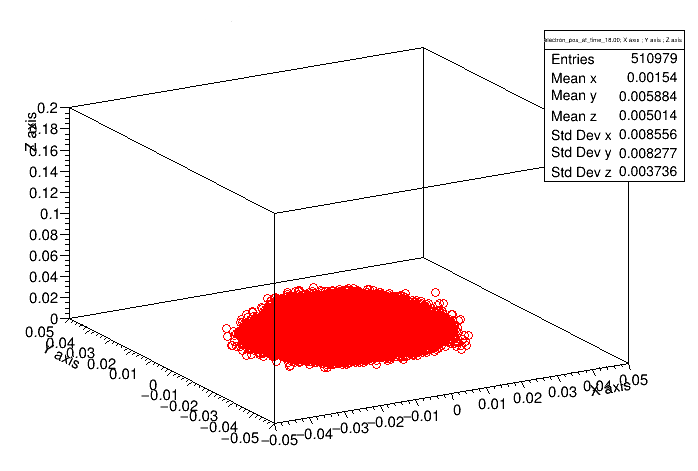}

}~~\subfloat[\label{fig:picofring}]{\includegraphics[width=0.27\textwidth]{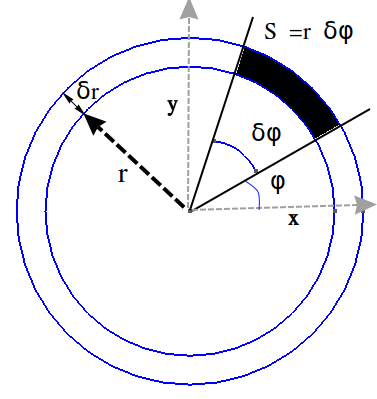}

}

\centering\subfloat[\label{fig:ringfieldpic}]{\includegraphics[width=0.35\textwidth]{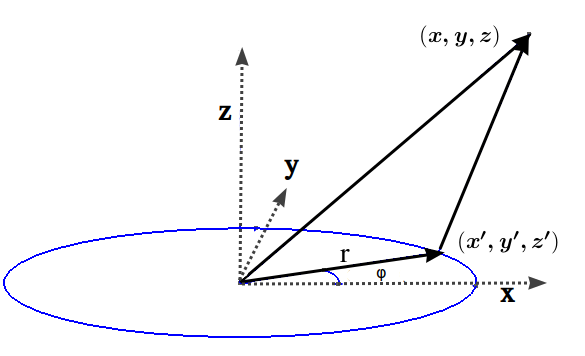}

}~~\subfloat[\label{fig:linefield}]{\includegraphics[width=0.40\textwidth]{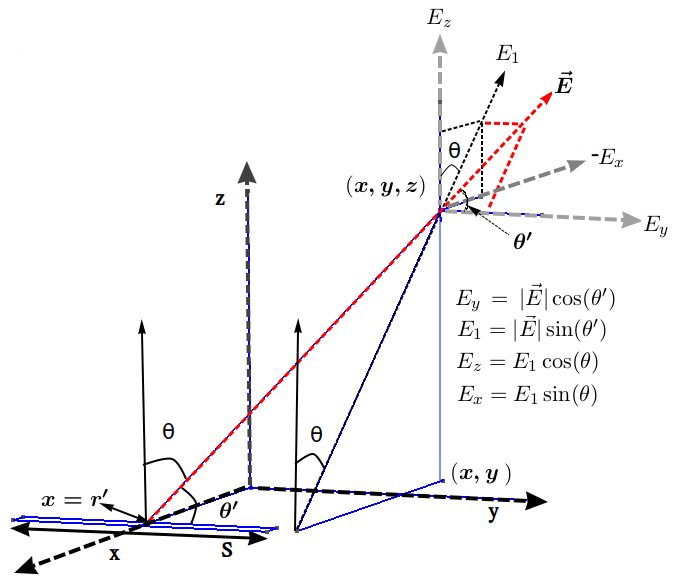}

}

\caption{(a) The position of electron clouds at time 18 ns where the total number
of electron is 510979. (b) picture of a ring of width $\delta r$
(c) Computation of electric field due to charged ring. (d) Components
of the electric field due to line charge.}
\end{figure}

An avalanche has been simulated inside an RPC
of dimension 30 cm $\times$ 30 cm and a 2 mm gas gap, from an electron
created at the origin (0,0,0) using the Garfield++ simulation tool \cite{Garfield++}.
The gas mixture containing 97\% of $C_{2}H_{2}F_{4}$, 2.5\% of $i-C_{4}H_{10}$
and 0.5\% of $SF_{6}$ has been selected. A uniform electric field of $50kV/cm$
is applied perpendicularly to the parallel plates of the detector
(which is considered here as z-direction) to perform this simulation.
The operating pressure of the gas has been kept equal to one atmospheric
pressure at temperature 293.15K. The Garfield++ can keep track of
the drifting position and time of each primary and secondary electrons
and ion generated during the avalanche. A table of position and corresponding time of those electrons and ions has been formed to calculate
interpolated positions of electrons at a certain instant of time.
The interpolated positions of space charge cloud at 18 ns, has been shown
in figure \ref{fig:position of elc clouds}.

\section{Calculation of the electric field due to the space charge distribution}

\label{subsec:algorithm_ring} 

\subsection{Ring approximation (method A) \label{subsubsec:ring-approximation}}

It is assumed that the avalanche charge distribution has a rotational symmetry about the z-axis as apparent from figure \ref{fig:position of elc clouds}.
Along the z-direction, the gas gap $g$ can be divided into $N_{z}$
steps with the step size $\delta z=\frac{g}{N_{z}}$. The space charge
region along the X-Y plane can also be divided into a number ($N_{r}$)
of concentric charged rings centered at z-axis and of gradually increasing
radius $r$ (see figure \ref{fig:picofring}). The size of the ring $\delta r=\frac{r_{max_{}}}{N_{r}}$
and $\delta z$ have been chosen according to the transverse and longitudinal
spread of the avalanche, e.g. $\delta r=\delta z=0.001\,cm$ and $r_{max}=0.045\,cm$.
Now, starting from a height $z=\bar{z}$ ( $\bar{z}<g$), if any charge
is present then the electric field due to all $N_{r}$ rings has been
calculated by doing numerical integration of the equation (\ref{eq:ring_field})
and (\ref{eq:p.fonte_cyl}) and then the value of $z$ set to $z+\delta z$.
This process is iterated till $z_{max}=g$. In this method,
the whole region of space charge can be covered to calculate the electric
field. The calculated components of the field due to all rings are summed to get the total
electric field at any point.

\subsubsection{Components of Electric field vector due to ring\label{subsec:Components-of-Electric}}

The X,Y and Z components of electric field at any position (x,y,z)
due to a ring of radius \textbf{r} and of uniform charged density $\lambda$ (see figure \ref{fig:ringfieldpic}),
centered at z-axis can be written as follows,

\begin{equation}
\left.\begin{gathered}\begin{array}{c}
E_{x}^{A}=\frac{\lambda\,\boldsymbol{r}}{4\pi\epsilon_{0}}\int\limits _{0}^{2\pi}\frac{(x-\boldsymbol{r}\cos\left(\phi\right))}{\Delta}d\phi,\,E_{y}^{A}=\frac{\lambda\,\boldsymbol{r}}{4\pi\epsilon_{0}}\intop\limits _{0}^{2\pi}\frac{(x-\boldsymbol{r}\sin(\phi))}{\Delta}d\phi\\
E_{z}^{A}=\frac{\lambda\,\boldsymbol{r}}{4\pi\epsilon_{0}}\intop\limits _{0}^{2\pi}\frac{\,\,\,\,(\,z-\bar{z}\,)\,\,\,}{\,\Delta\,}d\phi
\end{array}\end{gathered}
\right\} \label{eq:ring_field}
\end{equation}

where, $\Delta=[(x-r\cos(\phi))^{2}+(y-r\sin(\phi))^{2}+(z-\bar{z})^{2}]^{\frac{3}{2}}$, $\bar{z}$ is the position of center of that ring along z-axis, and $\phi$ is the angular displacement of an element of ring of length \textbf{r}$\delta\phi$ from the x-axis (see figure \ref{fig:picofring}).
If $Q_{ring}$ is total charge of that ring then $\lambda=\frac{Q_{ring}}{2\pi\boldsymbol{r}}$.

\subsection{Straight-line approximation with uniform charge density (method B)\label{subsec:Straight-line-approximation-with}}

A ring can be equally segmented into a number of straight-lines. If
$r$ and $\delta r$ are the radius and thickness of
that ring respectively, then the length of an arc of any segmented
element of that ring is S=$r\,\delta\phi$,
where $\delta\phi$ is the angle in radian, subtended to the center
of the circle (see figure \ref{fig:picofring}). $S$ can
be approximated as a straight-line of length $S$ and thickness $\delta r$,
when $\delta\phi$ is very small i.e. $S/r\leq1$.
It is discussed in section \ref{subsubsec:ring-approximation} that
the space charge region can be divided into a number of rings. As
an extension of this algorithm, those rings are also split into a
number of straight lines, where  $\delta\phi$ is chosen by the user. Therefore, the electric field
can be calculated for each charged straight-line (for case-1 and case-2
in section \ref{sec:Results-and-discussions} ) corresponding to a
ring using the equations (\ref{eq:stline_eqn}). Thus, after the formation
of a ring by a number of straight lines the same iteration process
discussed in section \ref{subsubsec:ring-approximation} can be followed
to get electric fields for all charges. The components of fields of
each charged line are added iteratively to get the total electric
field. Thus we can reproduce the results of the electric field of
charged rings using the line charge approximation.

\subsubsection{Components of Electric field vector due to a charged line\label{subsec:Components-of-Electric-1}}

Let us consider a straight line of constant charged density $\bar{\lambda}$
and of length $S$, aligned parallel to the y-axis at $x=\bar{r}$,
and $z=\bar{z}$. The X, Y, Z components of the electric
field at any position (x,y,z) due to this straight line can be written
as follows (see figure \ref{fig:linefield}),

\begin{equation}
\left.\begin{gathered}\begin{array}{c}
E_{x}^{B}=\frac{\bar{\lambda}(x-\bar{r})}{4\pi\epsilon_{0}P^{2}}\left[\frac{(y+\frac{S}{2})}{\sqrt{(y+\frac{S}{2})^{2}+P^{2}}}-\frac{(y-\frac{S}{2})}{\sqrt{(y-\frac{S}{2})^{2}+P^{2}}}\right],E_{y}^{B}=-\frac{\bar{\lambda}}{4\pi\epsilon_{0}}\left[\frac{1}{\sqrt{(y+\frac{S}{2})^{2}+P^{2}}}-\frac{1}{\sqrt{(y-\frac{S}{2})^{2}+P^{2}}}\right]\\
\\
E_{z}^{B}=\frac{\bar{\lambda}(z-\bar{z})}{4\pi\epsilon_{0}P^{2}}\left[\frac{(y+\frac{S}{2})}{\sqrt{(y+\frac{S}{2})^{2}+P^{2}}}-\frac{(y-\frac{S}{2})}{\sqrt{(y-\frac{S}{2})^{2}+P^{2}}}\right]
\end{array}\end{gathered}
\right\} \label{eq:stline_eqn}
\end{equation}

where, $P=\sqrt{(z-\bar{z})^{2}+(x-\bar{r})^{2}}$, and if $Q_{st}$
is the total charge of this straight line then, $\bar{\lambda}=\frac{Q_{st}}{S}$.

For a chosen value of $\delta\phi$ it can be said that the total number of straight-line needs to form a ring is $N_{st}$=$\frac{360}{\delta\phi}$. So $\delta\phi$ is
the minimum angle that has to rotate to reach from one segment to
another nearest segment. Necessary coordinate transformations have been carried out to evaluate electric field in a consistent frame of reference. 

\subsection{Method available in literature (method C)\label{subsec:Method-available-in}}

The equation of the electric field at any point ($\rho,\alpha,z$) in cylindrical co-ordinate system due to a ring of uniform charged
density $\lambda$ and radius "$a$" centered at z-axis, can also be
found in ref. \cite{book-electro}(v. I pp. 176), (see figure \ref{fig:ringfieldpic}),

\begin{equation}
\left.\begin{aligned}\vec{E}_{ring}^{C}(\rho,z,a) & =\frac{\lambda a}{\pi\epsilon_{0}}\left[\frac{1}{2\dot{r}_{1}\rho}\left(K_{1}(u)-\frac{(a^{2}-\rho^{2}+z^{2})K_{2}(u)}{\dot{r}_{1}^{2}(1-u^{2})}\right)\hat{\rho}+\frac{zK_{2}(u)}{\dot{r}_{1}^{3}(1-u^{2})}\hat{z}\right]=E_{r}^{C}\hat{\rho}+E_{z}^{C}\hat{z}\\
u & =\frac{2\sqrt{a\rho}}{\dot{r}_{1}},\,\dot{r}_{1}=\sqrt{(a+\rho)^{2}+z^{2}}
\end{aligned}
\right\} \label{eq:p.fonte_cyl}
\end{equation}

where $E_{r}^{C},E_{z}^{C}$ are the radial and z- components of electric
field and $K_{1}(u)$ and $K_{2}(u)$ are the complete elliptic integrals
of first and second kinds.

The components of fields calculated in Cartesian co-ordinates (for
method A and B) have been converted into cylindrical co-ordinates by
using ``Jacobi transformation'' to compare with the results of method
C. The required Jacobi matrix for this transformations is given below,

\begin{equation}
\begin{array}{c}
\begin{split}\left(\begin{array}{c}
E_{r}^{i}\\
E_{{\alpha}}^{i}\\
E_{z}^{i}
\end{array}\right) & =\left(\begin{array}{ccc}
\cos(\text{\ensuremath{{\alpha}}}) & \sin({\alpha}) & 0\\
-\sin({\alpha}) & \cos({\alpha}) & 0\\
0 & 0 & 1
\end{array}\right)\left(\begin{array}{c}
E_{x}^{i}\\
E_{y}^{i}\\
E_{z}^{i}
\end{array}\right), & {\alpha}=\tan^{-1}(\frac{y}{x})\end{split}
\end{array}
\end{equation}

$E_{r}^{i},E_{{\alpha}}^{i},E_{z}^{i}$ are the radial,
${\alpha}$ and z directional components of electric fields at a position ($\rho,\alpha,z$) of any
charged ring or line. Where, $x=\rho \cos{(\alpha)}$, $y=\rho \sin{(\alpha)}$ and z=z and $i=A$,$B$ for method-A and method-B respectively.
These cylindrical form of the components for different methods can
be represented together as $E_{r}^{A,B,C},E_{{\alpha}}^{A,B},E_{z}^{A,B,C}$.

\section{Results and discussions \label{sec:Results-and-discussions}}

\begin{figure}
\centering\includegraphics[scale=0.30]{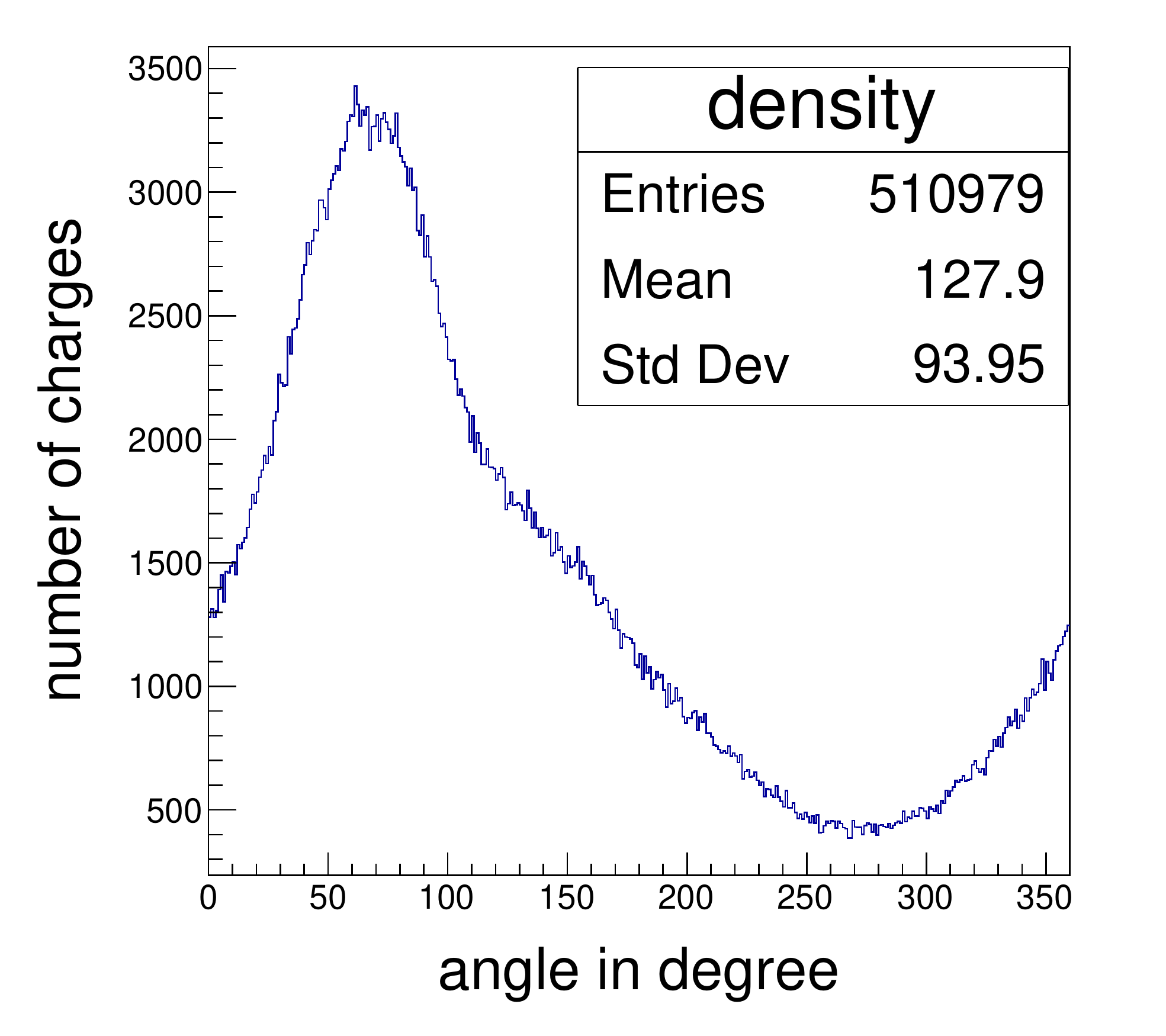}

\caption{\label{fig:angular-distribution-of-full-figure}Angular distribution
of charges at time 18 ns}
\end{figure}

The integrations of equations (\ref{eq:ring_field}) and (\ref{eq:p.fonte_cyl})
have been solved numerically by using standard "GSL-Integrator" from  "GSL-library" available in "root 6.18/04" \cite{root-cern,BrunRoot}.
The absolute values of calculated electric field components are set
to zero when it goes below a minimum number $\epsilon=10^{-8}$ ,
because at this range the field due to the space charge is negligible
in comparison to the applied field. The results can be divided into two cases\textbf{:} 
\begin{description}
\item [{case-1:}] The charge density $\lambda$ has been considered constant
in equations (\ref{eq:ring_field}) and (\ref{eq:p.fonte_cyl}) of methods
A,C. In method-B the charge ($Q_{ring}$) of a ring is equally shared
between segmented lines from that ring, which is $Q_{st}=\lambda S=\frac{Q_{ring}S}{2\pi r}$
for a line. 
\item [{case-2:}] In this case, the conditions for equations (\ref{eq:ring_field})
and (\ref{eq:p.fonte_cyl}) remain the same as in the above \textbf{case-1}.
However, in the actual scenario the angular distribution of the total
$Q_{ring}$ charge may not be uniform over the ring. Therefore, the
amount of charge $Q_{st}$ will be different for different $N_{st}$
line of that ring. So charges are calculated separately for each line.
Distribution of the charges reside on each line at different angles
for all radius and height, have been shown in figure \ref{fig:angular-distribution-of-full-figure}. 
\end{description}

\subsection{Results of case-1\label{subsec:Results-of-case-1}}

\begin{figure}
\subfloat[\label{fig:Ez-uniform-x0-y0}]{\includegraphics[scale=0.26]{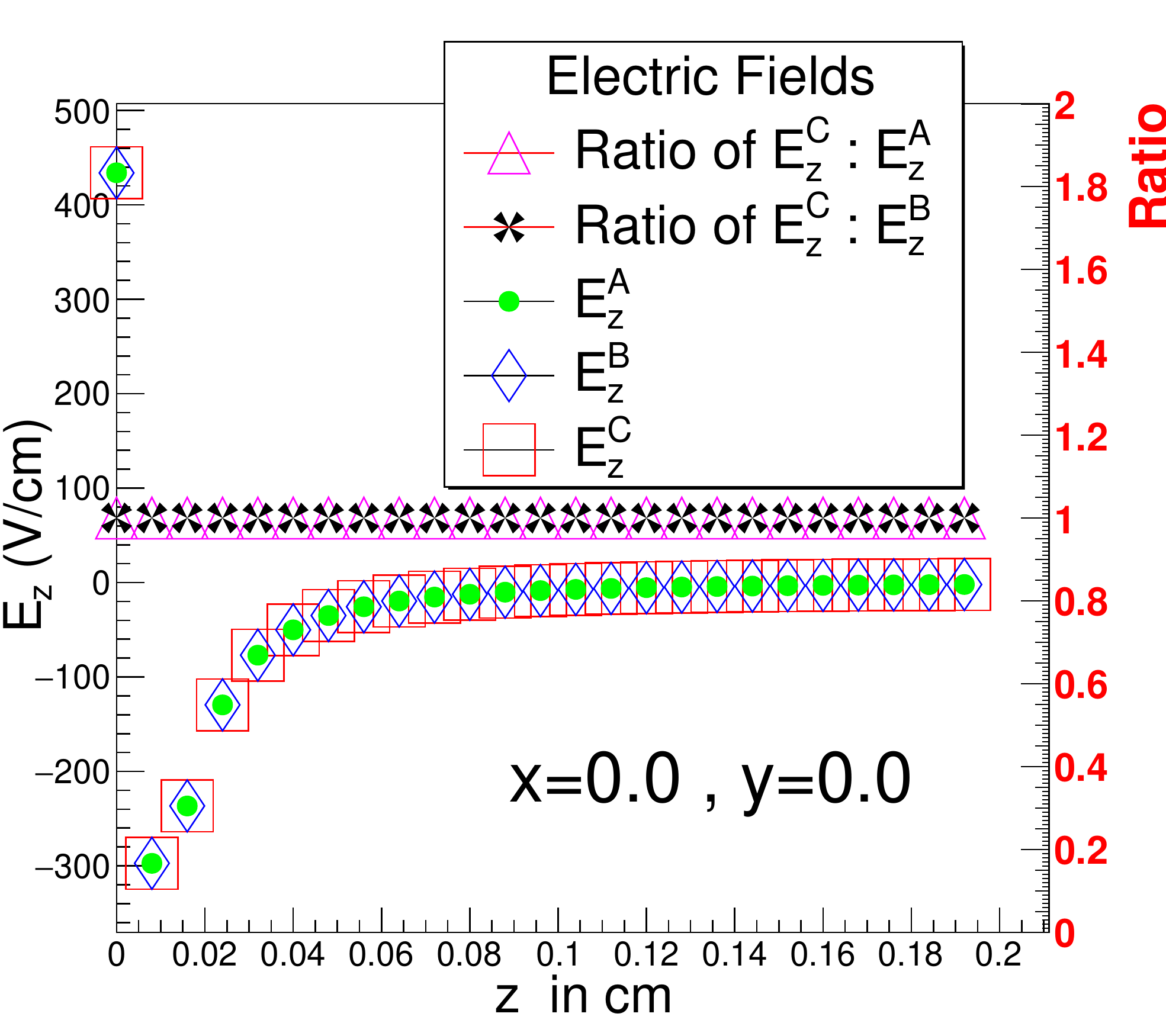}

}\subfloat[\label{fig:Er-uniform-x0-y0}]{\includegraphics[scale=0.26]{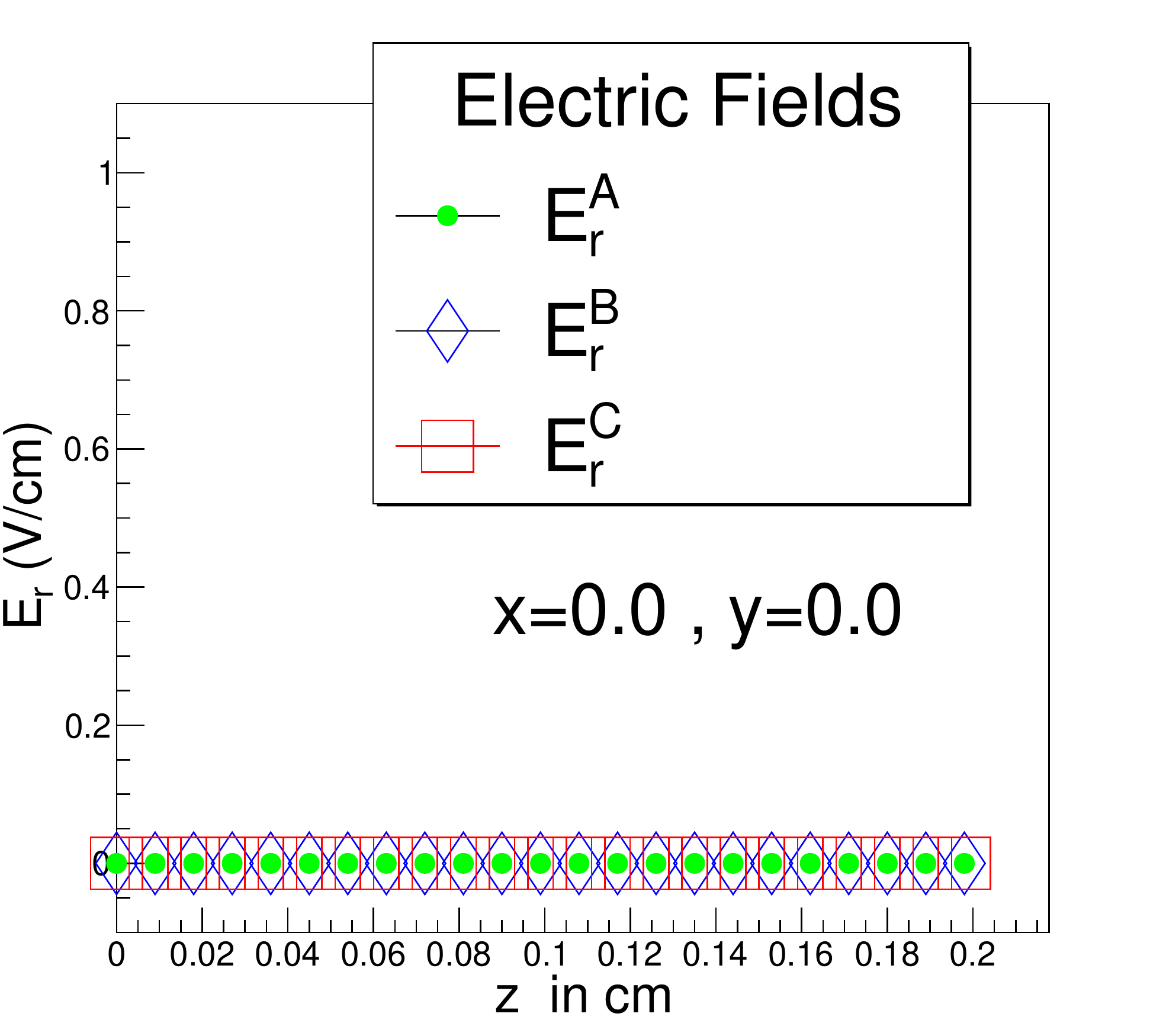}
}\subfloat[\label{fig:Ephi-uniform-x0-y0}]{\includegraphics[scale=0.22]{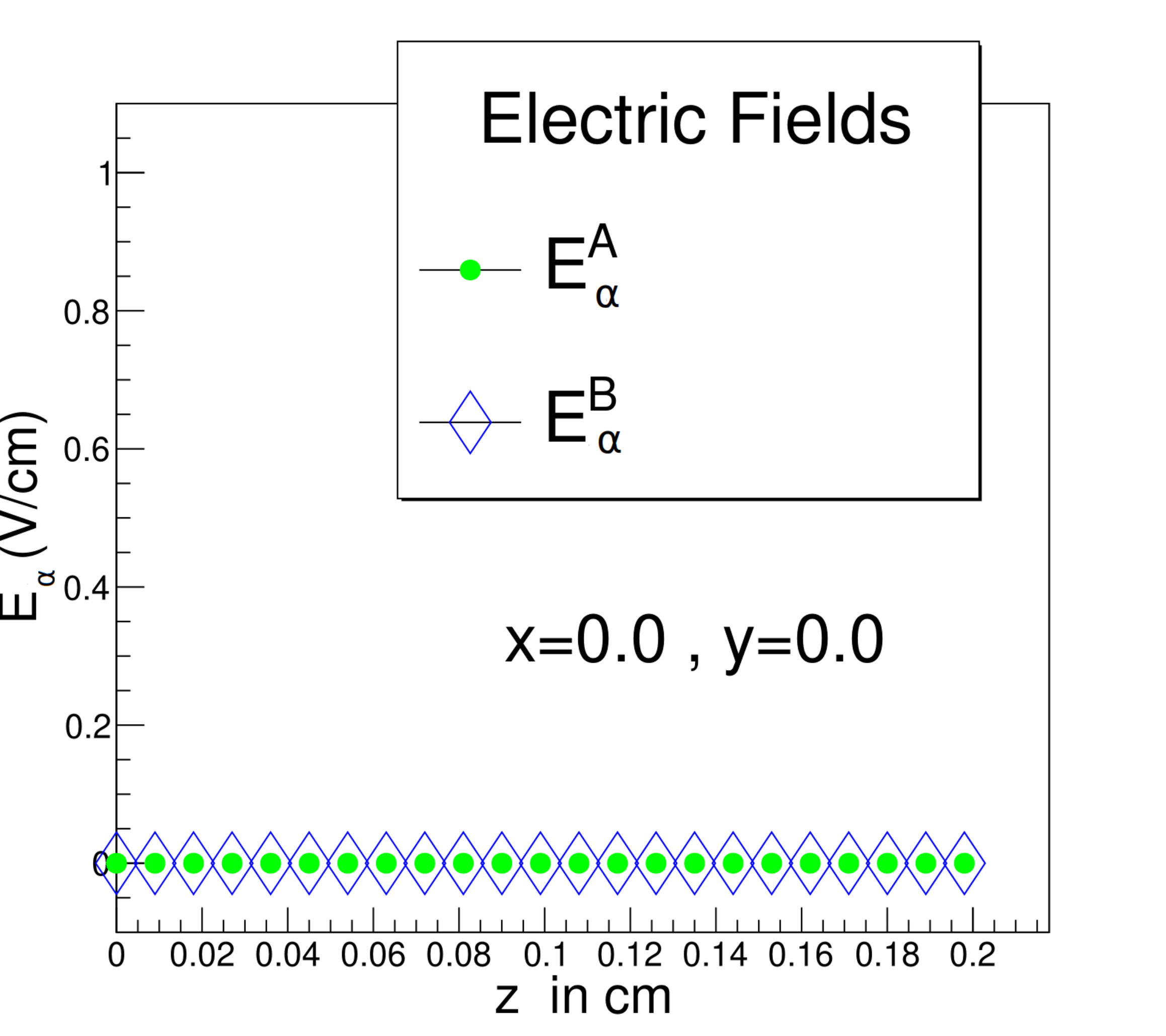}
}
\caption{\label{fig:uniform_density} Computation of electric field components
for case-1(section \ref{subsec:Results-of-case-1}) on the z-axis. (a) The
variation of $E_{z}^{A,B,C}$ on the z-axis has shown in the left-side
axis of the figures and right side axis contain the corresponding
ratios of the field components between $E_{z}^{C}$ and $E_{z}^{A,B}$,
(b) variation of $E_{r}^{A,B,C}$ on the z axis, (c) variation of
$E_{\alpha}^{A,B}$ on the z axis.}
\end{figure}

It is well known to us that the electric field component $E_{z}$
is dominating over $E_{r}$ and $E_{{\alpha}}$ on the z-axis of
a uniformly charged ring. Because of the axial symmetry, the components
$E_{r}$ and $E_{{\alpha}}$ cancel out each other and become zero.
It can also be verified from the figures \ref{fig:Ez-uniform-x0-y0},
\ref{fig:Er-uniform-x0-y0} and \ref{fig:Ephi-uniform-x0-y0}, where the
components of electric field $E_{r}^{A,B,C},E_{{\alpha}}^{A,B},E_{z}^{A,B,C}$
have been plotted for three different methods A,B,C. Again, from the
same figure \ref{fig:Ez-uniform-x0-y0} it is clear that the ratios $E_{z}^{C}/E_{z}^{A}=E_{z}^{C}/E_{z}^{B}=1$
on the z-axis. However, The values of $E_{r}^{A,B,C},E_{{\alpha}}^{A,B}$
are always zero on the z-axis, so the calculation of their ratios
is not possible . Hence, the ratio plot is not shown in the figures \ref{fig:Er-uniform-x0-y0}
and \ref{fig:Ephi-uniform-x0-y0}. 


\subsection{Results of case-2\label{subsec:Results-of-case-2}}

\begin{figure}
\subfloat[\label{fig:Ez-x=00003D00003D00003D0-y=00003D00003D00003D0-non-uni}]{\includegraphics[scale=0.263]{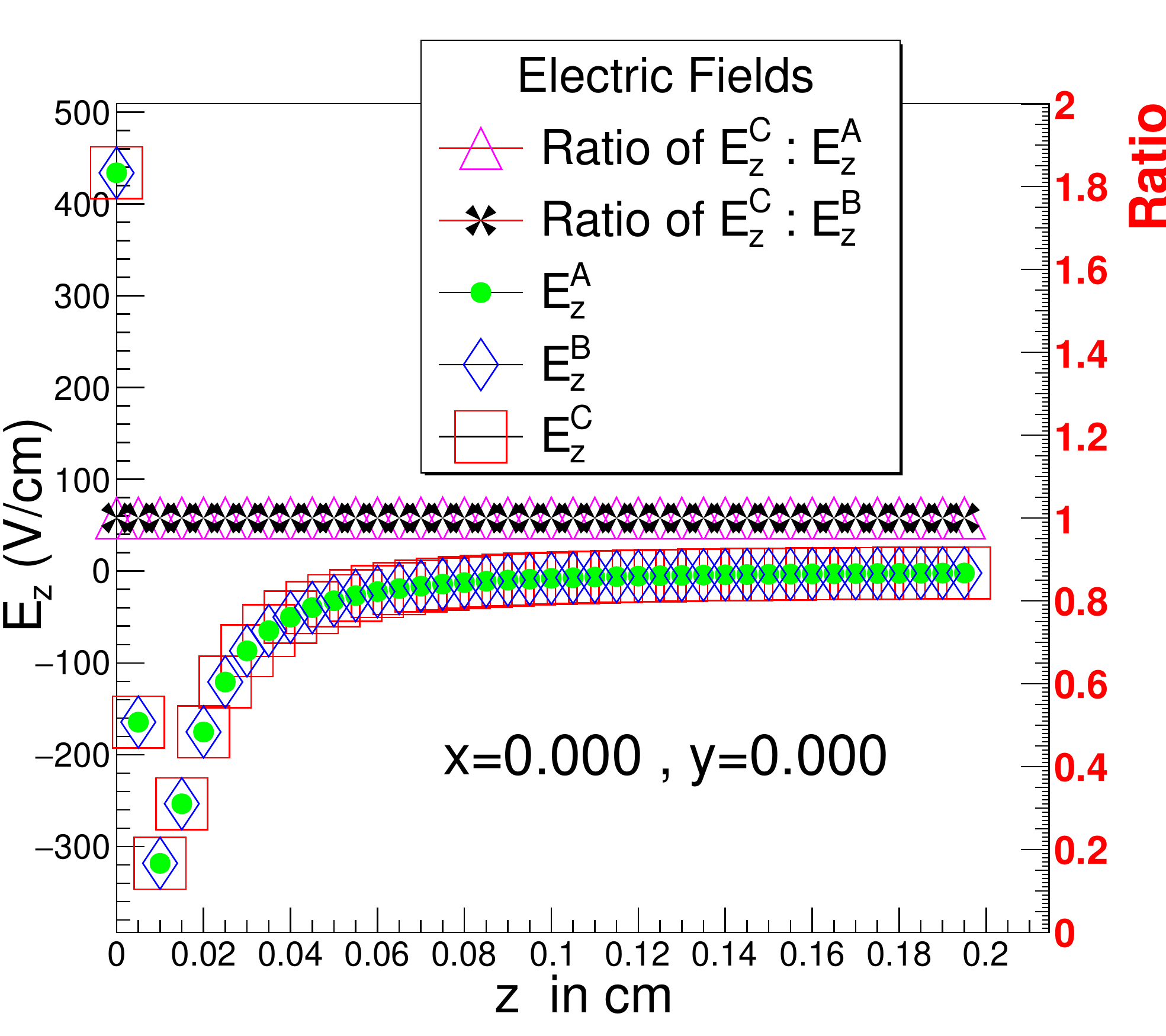}
}\subfloat[\label{fig:Er-x=00003D00003D00003D0-y=00003D00003D00003D0-non-uni}]{\includegraphics[scale=0.263]{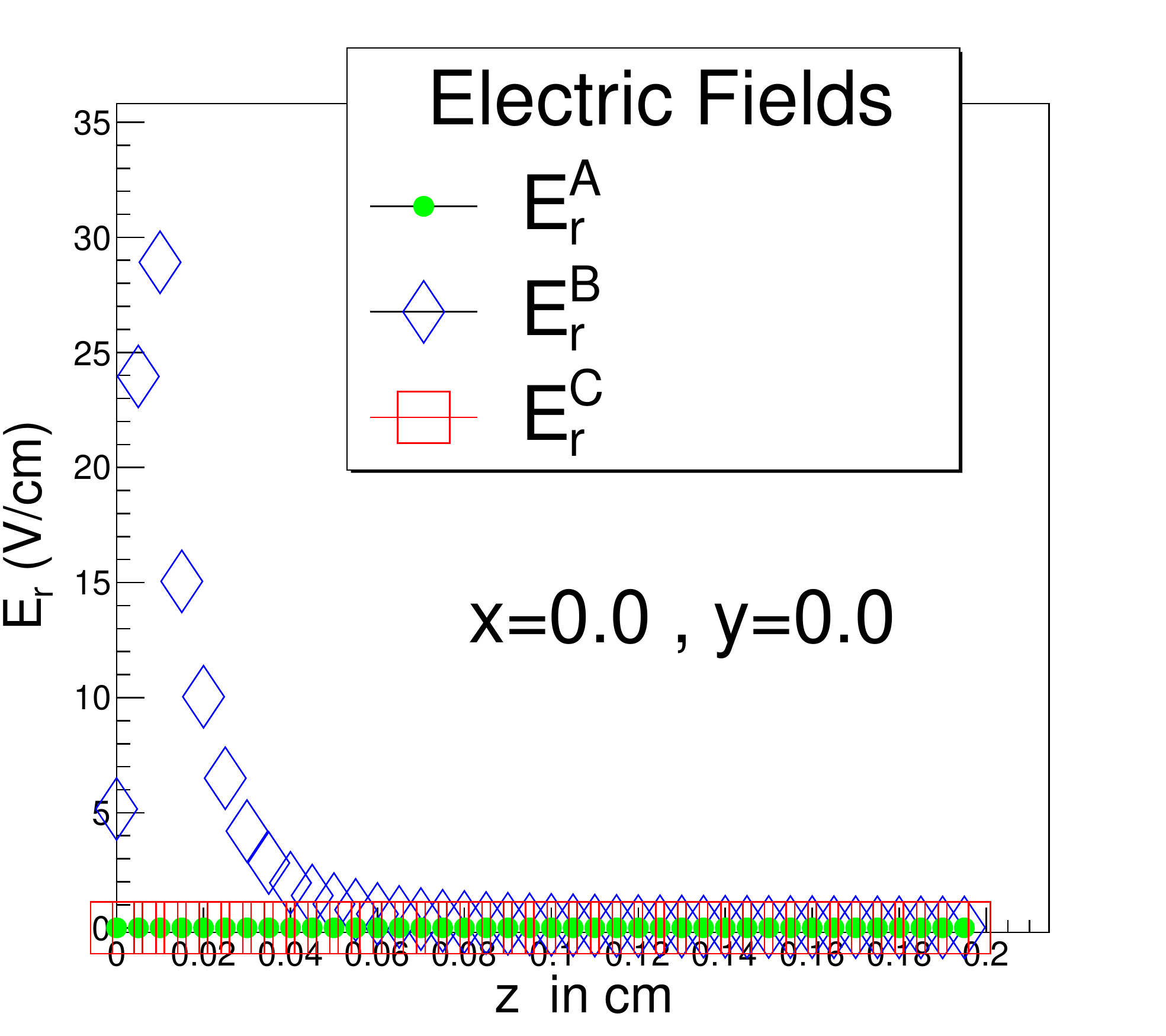}
}\subfloat[\label{fig:Ephi-x=00003D00003D00003D0-y=00003D00003D00003D0-non-uni}]{\includegraphics[scale=0.223]{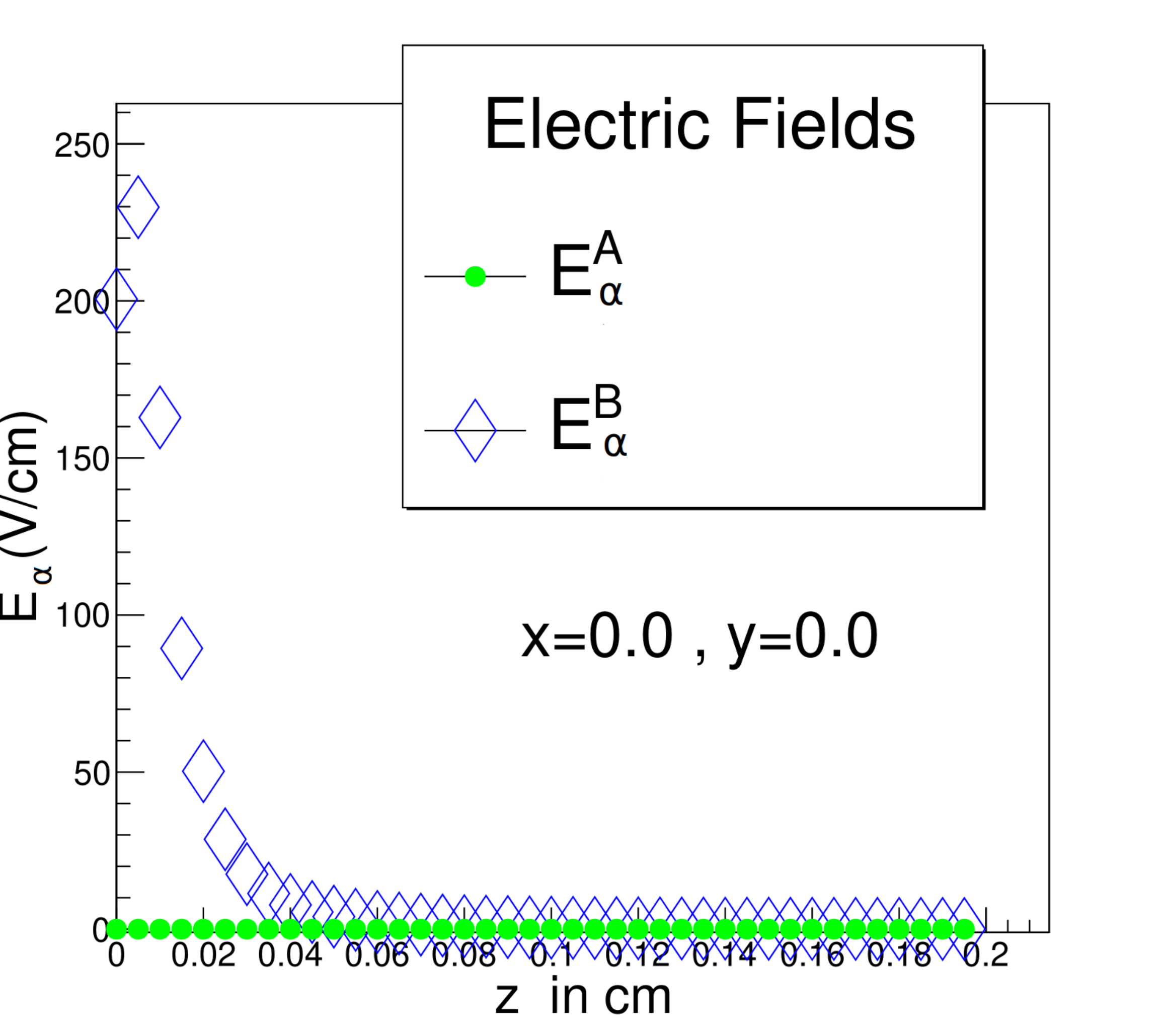}
}

\caption{\label{fig:1-1} Computation of electric field components for case-2(section \ref{subsec:Results-of-case-2})
on the z-axis. (a) The variation of $E_{z}^{A,B,C}$ on the z-axis
has shown in the left-side axis of the figures and right side axis
contain the corresponding ratios of the field components between $E_{z}^{C}$and
$E_{z}^{A,B}$, (b) variation of $E_{r}^{A,B,C}$ on the z axis, (c)
variation of $E_{\alpha}^{A,B}$ on the z axis.}
\end{figure}

The magnitudes of $E_{z}$ components for three methods A,B,C ($E_{z}^{A,B,C}$)
are the same which can be verified from the figure \ref{fig:Ez-x=00003D00003D00003D0-y=00003D00003D00003D0-non-uni}.
But there are discrepancies between the radial and ${\alpha}$ components of electric field
calculated in method-B ($E_{r}^{B},E_{{\alpha}}^{B}$) and method-A,C
( $E_{r}^{A,C},E_{{\alpha}}^{A,C}$) which is shown in the figures \ref{fig:Er-x=00003D00003D00003D0-y=00003D00003D00003D0-non-uni}
and \ref{fig:Ephi-x=00003D00003D00003D0-y=00003D00003D00003D0-non-uni}.
The components $E_{r}^{A,C},E_{{\alpha}}^{A,C}$ still gives the
same result zero along the z-axis as in case-1 for method-A and C. However,
the components $E_{r}^{B},E_{{\alpha}}^{B}$ is showing a non zero
value. Especially near the charge distribution, the value is much
higher than zero. These discrepancies can be explained from the angular
distribution of charges shown in figure \ref{fig:angular-distribution-of-full-figure}.
It is clear from the same figure \ref{fig:angular-distribution-of-full-figure}
that the nature of this angular distribution is not uniform; instead,
it is observed that most of the charges are within the angular range from $0$ to $150$ degrees.

\section{Conclusions}

Initially, it was assumed that the nature of the electron cloud has
some axial symmetry about the z-axis. So due to this symmetry, one
can easily neglect the $E_{{\alpha}}$ component. Subsequently, from the analysis of an avalanche from a single primary electron it is observed that the angular charge distribution is not uniform, so the component $E_{{\alpha}}$
plays a significant role while avalanche is developing. Hence, it
can't be ignored anymore. It is already discussed that the avalanche is generated here from a single primary electron. However, in an actual event, the avalanche
can be formed from several primaries. So the results for that
case along with the reflections of charges on the ground plates  need to be found, which is also our future interest.

The merit of using straight line-approximation is that it produces
similar results as uniformly charged rings as well as it can be easily
used when the charged density is nonuniform over the ring. Again the
most remarkable feature of the same is the components of field equations
do not contain any elliptical integrals. So we do not need to worry
about the numerical integrations.

\appendix

\section{Acknowledgement}

I am grateful to the members of INO collaboration and HEP experiment
division of VECC for providing me the opportunity to do research.
I would like to thank Prof. Nayana Majumdar for intense discussion
and valuable suggestions. 
\bibliographystyle{JHEP.bst}
\bibliography{numerical_space_charge_No_model_lit}

\providecommand{\href}[2]{#2}\begingroup\raggedright\begin{thebibliography}{1}

\bibitem{cardeli-1}
R.~Santonico and R.~Cardarelli, \emph{Development of resistive plate counters},
  \href{http://dx.doi.org/https://doi.org/10.1016/0029-554X(81)90363-3}{\emph{Nuclear
  Instruments and Methods in Physics Research} {\bfseries 187} (1981)
  377--380}.

\bibitem{cardeli-2}
R.~Cardarelli, R.~Santonico, A.~Biagio and A.~Lucci, \emph{Progress in
  resistive plate counters},
  \href{http://dx.doi.org/https://doi.org/10.1016/0168-9002(88)91011-X}{\emph{Nuclear
  Instruments and Methods in Physics Research Section A: Accelerators,
  Spectrometers, Detectors and Associated Equipment} {\bfseries 263} (1988) 20
  -- 25}.

\bibitem{Sauli:2014cyf}
F.~Sauli, \emph{{Gaseous Radiation Detectors}: {Fundamentals and
  Applications}}, vol.~36.
\newblock Cambridge University Press, 8, 2014,
  \href{http://dx.doi.org/10.1017/CBO9781107337701}{10.1017/CBO9781107337701}.

\bibitem{Lippmann_1}
C.~Lippmann and W.~Riegler, \emph{{Space charge effects in resistive plate
  chambers}}, \href{http://dx.doi.org/10.1016/j.nima.2003.08.174}{\emph{Nucl.
  Instrum. Meth. A} {\bfseries 517} (2004) 54--76}.

\bibitem{Garfield++}
H.~Schindler, \emph{Garfield++ user$'$s guide},
  {\emph{\href{https://garfieldpp.web.cern.ch/garfieldpp/documentation/UserGuide.pdf}{https://garfieldpp.web.cern.ch/garfieldpp}}
  (April, 2020) }.

\bibitem{book-electro}
E.~Durand, \emph{$\acute{E}lectrostatique$}.
\newblock Masson et $C^{ie}$, 1964.

\bibitem{root-cern}
\emph{Root data analysis framework user$'$s guide},
  {\emph{\href{https://root.cern.ch/root/htmldoc/guides/users-guide/ROOTUsersGuideA4.pdf}{https://root.cern.ch/root/htmldoc/guides/users-guide/ROOTUsersGuideA4.pdf}}
  (May, 2018) }.

\bibitem{BrunRoot}
R.~Brun and F.~Rademakers, \emph{{ROOT: An object oriented data analysis
  framework}},
  \href{http://dx.doi.org/10.1016/S0168-9002(97)00048-X}{\emph{Nucl. Instrum.
  Meth. A} {\bfseries 389} (1997) 81--86}.

\end{thebibliography}\endgroup
\end{document}